# Know Your Author: Does the AI Penalty Hold in Short Fiction?

Michael Todasco

James Silberrad Center for Artificial Intelligence, San Diego State University

Joselyn Cesare

San Diego State University

Correspondence: mtodasco@sdsu.edu

## Abstract

Public concern about an “AI penalty” suggests that labeling content as AI-generated may negatively influence how it is evaluated. We tested this claim in a preregistered experiment (N = 254, per protocol) using a pure attribution design: participants read one of two ~200-word vignettes and were randomly assigned to see it labeled as Human-written, AI-written, or presented with no author line. Authorship labels did not produce reliable main effects on creativity, enjoyment, recommendation, or originality; observed effect sizes were uniformly small. However, labels strongly influenced inferred effort: participants estimated that Human-labeled stories took far longer to create than AI-labeled stories (back-transformed geometric means from ln[minutes + 1]: 148 vs. 6 minutes). Across conditions, higher inferred effort predicted greater enjoyment, and this relationship was also present within the AI-labeled condition. Additionally, participants’ prior attitudes toward AI moderated recommendation judgments: more positive attitudes were associated with higher recommendation ratings for AI-labeled stories, but not for Human-labeled stories. These findings suggest that while AI authorship labels do not systematically alter average evaluations of short fiction, they meaningfully shape perceptions of effort and interact with prior beliefs to influence downstream judgments.

## Introduction

As generative language models proliferate, creators and platforms face pressure to disclose whether a work was produced by a human author or an AI system. A growing literature suggests an "AI penalty": when people believe an artifact was created by AI, they may evaluate it less favorably than comparable human-created work, particularly in aesthetic domains (Bellaiche et al., 2023; Magni et al., 2024).

However, much of the evidence compares different human and AI artifacts, leaving open whether observed differences reflect authorship beliefs, differences in the artifacts themselves, or both. Pure attribution designs address this limitation by holding content constant while manipulating only the claimed source.

We apply a pure attribution design to short fiction. With fiction, evaluation is subjective, perceived effort matters, and readers typically care less about factual accuracy than in journalism or academic writing. Two brief vignettes were generated by a single AI system, then randomly labeled as Human-written, AI-written, or left Unlabeled. This design allows us to estimate the causal effect of an AI label on evaluations while controlling for the text.

We also test a plausible mechanism: the effort heuristic. People often infer quality from perceived effort, rewarding work that seems to require more time or labor (Kruger et al., 2004). AI labels may sharply reduce inferred effort ("generated in seconds"), which could reduce enjoyment or willingness to recommend even when the text is identical.

Finally, we examine whether attitudes toward AI moderate label effects on recommendation. Prior work suggests that beliefs about AI can shape downstream judgments even when quality is held constant (Lermann Henestrosa & Kimmerle, 2024).

## Hypotheses

H1 (AI penalty): Stories labeled as AI-written will receive lower ratings than stories labeled as Human-written on creativity, enjoyment, recommendation, and originality.

H2 (intermediate unlabeled): Stories with no authorship attribution will fall between the Human-labeled and AI-labeled conditions, though the unlabeled condition also removes explicit author framing entirely.

H3 (effort heuristic within AI): Within the AI-labeled condition, higher perceived author effort (operationalized as perceived creation time) will be associated with higher enjoyment.

# Method

## Participants

We recruited 300 US-based participants via Prolific. Compensation was $2.00; median completion time was 5 min 9 s. Preregistered exclusion criteria removed participants who failed either the authorship-recall manipulation check or the story-comprehension attention question. A total of 46 participants were excluded: 43 failed the authorship-recall check, and 4 failed the comprehension check (1 participant failed both), yielding a final per-protocol sample of N = 254.

## Data Preparation

Participants estimated how long the vignette took to create using an open-ended response. We converted responses to minutes when they contained sufficient information to quantify duration (e.g., numeric values with or without units such as “10,” “10 minutes,” or “about an hour”). Responses that were missing or non-quantifiable (e.g., “not sure,” “a while”) were excluded from time-based analyses. This yielded N = 234 quantifiable time estimates from the N = 254 per-protocol sample. For all perceived-time analyses, we use this maximal quantifiable subset rather than a narrower formatting-based subset so that very short but still interpretable estimates remain in the data. Raw-minute correlations are reported for transparency, but inferential models use ln(minutes + 1) because the raw distribution is strongly right-skewed.

## Design

The study used a 3 (attribution label: AI vs Human vs Unlabeled) x 2 (story: Firefly vs Endless) between-subjects design. Participants read a single vignette preceded by an attribution line. In the Unlabeled condition, the entire author line was omitted.

## Materials

Anthropic’s Claude 3.7 Sonnet generated two ~200-word vignettes from the same prompt (“summer-appeal short story, creative ending”): Firefly Summer (May 13, 2025) and The Endless Summer (May 15, 2025). Full texts for the stories are available at https://doi.org/10.17605/OSF.IO/Z5KJG.

The attribution manipulation presented one of the following lines above the story: (a) Human: "Author: J. Taylor, an American short fiction writer." (b) AI: "Author: Claude, an AI system specialized in creative writing." (c) Unlabeled: no author information.

Both vignettes were AI-generated, so the Human-labeled condition involves deliberate misattribution. This allows us to isolate the label's causal effect.

## Measures

Primary outcomes. Participants rated the story's creativity, enjoyment, recommendation likelihood, and originality on 1-7 Likert scales. We also report an evaluation index computed as the mean of these four items (Cronbach's alpha = .918).

Perceived creation time. Participants provided an open-ended estimate of how long they believed the author spent writing and refining the story. Coding and transformation details are described in Data Preparation above.

AI attitudes. Participants reported AI familiarity and general positivity toward AI using single items on 1-5 scales. For moderation models, AI positivity was z-scored (mean 0, SD 1).

Manipulation checks. Authorship recall was assessed with a multiple-choice item asking who wrote the story. Story comprehension was assessed with a multiple-choice question tailored to the assigned vignette.

### Analytic approach

Confirmatory analyses followed the preregistration. We tested H1 and H2 using one-way ANOVAs to compare attribution conditions across the four preregistered evaluation outcomes. We assessed generalization across stories with two-way ANOVAs including story as a factor. H3 was tested as the association between perceived creation time and enjoyment within the AI-labeled condition, using the quantifiable time subset. Because perceived time was highly skewed, the primary effort-heuristic analyses use Pearson correlations and OLS with ln(minutes + 1) as the transformed predictor; raw-minute correlations are also reported in the appendix. Unless noted otherwise, tests are two-tailed with alpha = .05.

## Results

### Manipulation checks and attrition

Authorship recall was imperfect: 14.3% of participants (43/300) failed to identify the intended author label. Forty-two of those 43 passed the story-comprehension item, suggesting that the label was sometimes weakly encoded even when the vignette itself was understood. Recall failures were concentrated in the Human + Firefly cell (27% vs 12% elsewhere; Fisher's exact $p = .012$), producing modest cell-size imbalances in the per-protocol sample (Table 1).

### Primary outcomes

H1 and H2 were not supported. Attribution labels did not produce reliable differences in evaluations. Across the four preregistered outcomes, and the supplementary evaluation index, one-way ANOVAs showed no significant effect of attribution (all $ps > .22$; Table 3). For the evaluation index, mean ratings were Human $M = 4.72$ ($SD = 1.28$), AI $M = 4.52$ ($SD = 1.54$), and Unlabeled $M = 4.37$ ($SD = 1.40$).

Notably, the direction ran counter to H2: unlabeled stories received the lowest average evaluation index ($M = 4.37$), below both the AI-labeled ($M = 4.52$) and Human-labeled ($M = 4.72$) conditions. This ordering is the opposite of the predicted intermediate position, though the differences were not statistically significant.

These conclusions are unchanged under the preregistered Bonferroni correction for the four primary outcomes.

## Story robustness check

Two-way ANOVAs including story yielded no Condition × Story interactions for any outcome ($ps \geq .377$). There was a marginal main effect of story on enjoyment, favoring Firefly, $F(1, 248) = 3.17$, $p = .076$, but this effect did not interact with attribution.

## Perceived creation time

Attribution labels strongly shifted inferred effort. Among participants providing quantifiable time estimates ($N = 234$), perceived creation time differed by attribution on ln(minutes + 1), $F(2, 231) = 55.65$, $p < .001$, eta^2 = .325. Back-transforming log means to geometric means, Human-labeled stories were estimated to take 148 minutes, Unlabeled stories 75 minutes, and AI-labeled stories 6 minutes. Post hoc Tukey tests on ln(minutes + 1) showed that both Human and Unlabeled conditions estimated significantly more time than the AI condition ($ps < .001$). In contrast, Human vs Unlabeled was not significant ($p = .063$). In ratio terms, Human-labeled stories were estimated to take about 23 times longer than AI-labeled stories, and Unlabeled stories about 12 times longer.

## Exploratory story-specific enjoyment

As a supplementary follow-up, we examined whether the two fixed vignettes differed in enjoyment within label conditions. The omnibus Condition × Story interaction was not significant, $F(2, 248) = 0.98$, $p = .377$, so we do not interpret any single within-cell contrast as evidence of a reliable content × label interaction. Still, the largest story-specific difference appeared in the Human condition: Human–Firefly received higher enjoyment ratings than Human–Endless, $F(1, 75) = 4.45$, $p = .038$. In the quantifiable-time subset, Firefly also elicited higher perceived writing time than Endless within the Human condition, $F(1, 66) = 5.83$, $p = .019$, and controlling for perceived time attenuated the within-Human story difference in enjoyment. We treat this pattern as descriptive rather than confirmatory.

## Effort heuristic test (H3)

Supporting H3, perceived creation time was positively associated with enjoyment within the AI-labeled condition. Within AI-labeled participants with quantifiable time estimates ($N = 78$), ln(minutes + 1) correlated positively with enjoyment, $r = .37$, $p = .001$. The same positive association also appeared in the Human and Unlabeled groups; across all conditions ($N = 234$), the correlation was $r = .29$, $p < .001$ (Table 4b).

## Moderation by AI attitudes

Although attribution labels did not shift average recommendations, AI attitudes moderated willingness to recommend. In a regression predicting recommendation from attribution, story, AI positivity (z), and the Condition × AI-positivity interaction ($N = 253$), the interaction was significant, $F(2, 246) = 5.19$, $p = .006$ (Table 6). Simple slopes showed that AI positivity strongly predicted recommending AI-labeled stories ($b = 1.17$), but less so for Human-labeled stories ($b = 0.39$). Consequently, participants with low AI positivity preferred Human-labeled over AI-labeled stories, whereas this difference disappeared at high AI positivity.

### Exploratory mediation analyses

To probe mechanism, we estimated exploratory mediation models in which attribution (AI vs comparator) influenced enjoyment via perceived creation time (ln(minutes + 1)), estimated separately within each story using nonparametric bootstrap confidence intervals. Across both stories and contrasts (AI vs Human; AI vs Unlabeled), AI labeling predicted substantially lower perceived time, and perceived time predicted higher enjoyment, yielding negative indirect effects with 95% bootstrap CIs excluding zero (Table 7). Direct effects were small and mixed in sign, consistent with the near-zero total effects observed in the primary ANOVAs.

## Discussion

In this pure attribution design, explicit AI authorship labels did not reliably change readers' evaluations of short fiction when the text itself was held constant. H1 and H2 were not supported: across multiple rating dimensions, average differences between AI-labeled, Human-labeled, and Unlabeled conditions were small and statistically non-significant.

At the same time, attribution labels meaningfully shaped inferences about effort. Participants inferred dramatically less writing-and-refining time when the exact same story was labeled as AI-written. These effort inferences mattered: perceived creation time predicted enjoyment overall and within the AI-labeled condition, supporting the preregistered effort-heuristic hypothesis.

Attitudes toward AI further qualified how labels matter. Readers with more negative AI attitudes were less willing to recommend AI-labeled stories, while more positive readers showed little to no penalty. This moderation helps explain why average label effects were near zero: subgroups move in opposite directions.

A practical implication is that disclosure may influence downstream judgments less through a direct penalty and more through (a) whether the label is noticed and remembered and (b) what the label implies about effort. Notably, 14% of participants failed to recall the author label despite being shown it immediately before reading, suggesting that attribution can be weakly encoded in low-stakes contexts.

Limitations include the use of two short, uplifting vignettes generated by a single AI system, a U.S. Prolific sample, and single-item measures of AI attitudes and familiarity. With 254 participants across three conditions, the design was adequately powered to detect medium effects ($\eta^2 \approx .04$) but not the small effects observed here ($\eta^2 = .006$–$.012$).

Taken together, these results suggest that if an AI penalty exists in this setting, it is likely small. The observed pattern, in which unlabeled stories scored lowest, is also consistent with no penalty at all. Because the study was not optimized to detect small effects, larger replications with more precise measurements are needed to determine whether the true effect is a small penalty, effectively zero, or even a slight benefit of explicit attribution.

Additionally, the Unlabeled condition omitted the entire author line rather than presenting a neutral placeholder (e.g., 'Author: Anonymous'), so it differs from the labeled conditions in both attribution content and the presence of author framing. This confound limits the interpretation of H2 specifically, though it does not affect H1 or H3.

## Ethics

The study involved minor deception (authorship label) and concluded with a debrief. Participation risks were minimal; data were collected anonymously. The study was approved by the San Diego State University IRB (IRB# 25-0235).

## Future directions

The present study suggests explicit AI authorship labels do not diminish audience evaluations when text is held constant. However, the audience's judgments reflect perceived effort and general AI attitudes. Future research should address three priorities. First, generalize beyond two short vignettes by varying genre, length, quality, and author-label salience. The AI penalty may surface when credibility is central or text quality is low. Second, directly manipulate perceived effort to establish causality for both human and AI work.

Third, upgrade measurement precision using multi-item scales for AI attitudes and perceived effort rather than open-ended estimates. These findings need validation beyond US Prolific samples through cross-cultural and naturalistic replications.

## Data and materials availability

Anonymized data, materials, preregistration, and analysis code are available at the Open Science Framework (https://doi.org/10.17605/OSF.IO/Z5KJG).

## Appendix: Tables and Figures

**Table 1. Per-protocol sample size by condition and story.**

| Condition | Endless | Firefly |
|---|---|---|
| AI | 43 | 40 |
| Human | 43 | 34 |
| Unlabeled | 47 | 47 |

**Table 2. Outcome means (SD) by condition and story.**

| Condition | Story | N | Index | Creativity | Enjoyment | Recommend | Originality |
|---|---|---|---|---|---|---|---|
| AI | Endless | 43 | 4.60 (1.51) | 4.72 (1.55) | 4.88 (1.65) | 4.12 (1.92) | 4.70 (1.55) |
| AI | Firefly | 40 | 4.42 (1.58) | 4.50 (1.62) | 4.88 (1.42) | 3.83 (1.97) | 4.47 (1.71) |
| Human | Endless | 43 | 4.63 (1.40) | 4.77 (1.36) | 4.79 (1.39) | 4.26 (1.79) | 4.72 (1.49) |
| Human | Firefly | 34 | 4.84 (1.14) | 4.74 (1.26) | 5.44 (1.28) | 4.38 (1.48) | 4.79 (1.34) |
| Unlabeled | Endless | 47 | 4.32 (1.46) | 4.45 (1.52) | 4.62 (1.65) | 3.83 (1.77) | 4.38 (1.54) |
| Unlabeled | Firefly | 47 | 4.42 (1.35) | 4.47 (1.54) | 5.00 (1.49) | 3.85 (1.78) | 4.36 (1.51) |

**Table 3. One-way ANOVAs testing attribution effects on story evaluations (per-protocol N = 254).**

| Outcome | df1 | df2 | F | p | eta^2 |
|---|---|---|---|---|---|
| Index | 2 | 251 | 1.33 | .266 | .011 |
| Creativity | 2 | 251 | 0.85 | .427 | .007 |
| Enjoyment | 2 | 251 | 0.71 | .492 | .006 |
| Recommend | 2 | 251 | 1.52 | .221 | .012 |
| Originality | 2 | 251 | 1.35 | .261 | .011 |

**Table 4a. Correlations between perceived minutes and enjoyment (time-quantifiable subset, N = 234).**

| Group | N | r | p |
|---|---|---|---|
| Overall | 234 | .13 | .053 |
| AI | 78 | .23 | .046 |
| Human | 68 | .17 | .161 |
| Unlabeled | 88 | .14 | .185 |

**Table 4b. Correlations between ln(perceived minutes + 1) and enjoyment (time-quantifiable subset, N = 234).**

| Group | N | r | p |
|---|---|---|---|
| Overall | 234 | .29 | <.001 |
| AI | 78 | .37 | .001 |
| Human | 68 | .33 | .006 |
| Unlabeled | 88 | .33 | .002 |

**Table 5a. One-way ANOVA on ln(perceived minutes + 1) by condition (N = 234).**

| Outcome | df1 | df2 | F | p | eta^2 | omega^2 |
|---|---|---|---|---|---|---|
| ln(minutes + 1) | 2 | 231 | 55.65 | <.001 | .325 | .318 |

**Table 5b. Tukey HSD comparisons on ln(perceived minutes + 1) by condition (N = 234).**

| Group 1 | Group 2 | Mean diff (Group 2-1) | p | CI low | CI high | Reject |
|---|---|---|---|---|---|---|
| AI | Human | 3.01 | <.001 | 2.29 | 3.73 | True |
| AI | Unlabeled | 2.34 | <.001 | 1.66 | 3.01 | True |
| Human | Unlabeled | -0.67 | .063 | -1.37 | 0.03 | False |

**Table 5c. Descriptive statistics for perceived creation time by condition (quantifiable subset, N = 234).**

| Condition | N | Mean (min) | Median (min) | Geometric | Mean |
|---|---|---|---|---|---|

|  |  |  |  | mean (min) | ln(min+1) |
|---|---|---|---|---|---|
| AI | 78 | 20.31 | 5.00 | 6.36 | 2.00 |
| Human | 68 | 1131.91 | 120.00 | 148.13 | 5.00 |
| Unlabeled | 88 | 871.56 | 60.00 | 75.17 | 4.33 |

**Table 6. Moderation of recommendation by AI positivity (N = 253).**

| Test | df1 | df2 | F | p | eta_p^2 |
|---|---|---|---|---|---|
| Condition x AI positivity (Type-II ANOVA) | 2 | 246 | 5.19 | .006 | .040 |
| Condition x AI positivity (HC3 joint Wald) | 2 | 246 | 5.16 | .006 | - |

Note. N = 253 because one participant provided a nonnumeric AI-positivity response ("No Opinion") and was therefore excluded from this model.

**Table 7. Story-specific mediation of AI label effects on enjoyment via perceived creation time (quantifiable subset; 5,000 bootstrap samples).**

| Story | Contrast | N | a | b | c' | ab | CI low | CI high |
|---|---|---|---|---|---|---|---|---|
| Endless | AI vs Human | 79 | -2.476 | .401 | 1.106 | -0.993 | -1.597 | -0.508 |
| Endless | AI vs Unlabeled | 85 | -1.973 | .376 | 1.076 | -0.742 | -1.249 | -0.351 |
| Firefly | AI vs Human | 67 | -3.675 | .165 | .053 | -0.608 | -1.198 | -0.093 |
| Firefly | AI vs Unlabeled | 81 | -2.706 | .207 | .528 | -0.560 | -1.036 | -0.117 |

Note. a = effect of label on ln(minutes + 1); b = effect of ln(minutes + 1) on enjoyment; c' = direct effect of label on enjoyment; ab = indirect effect. 95% CIs are bias-corrected bootstrap.

**Figure 1. Evaluation index by attribution label (mean ± 95% CI).**

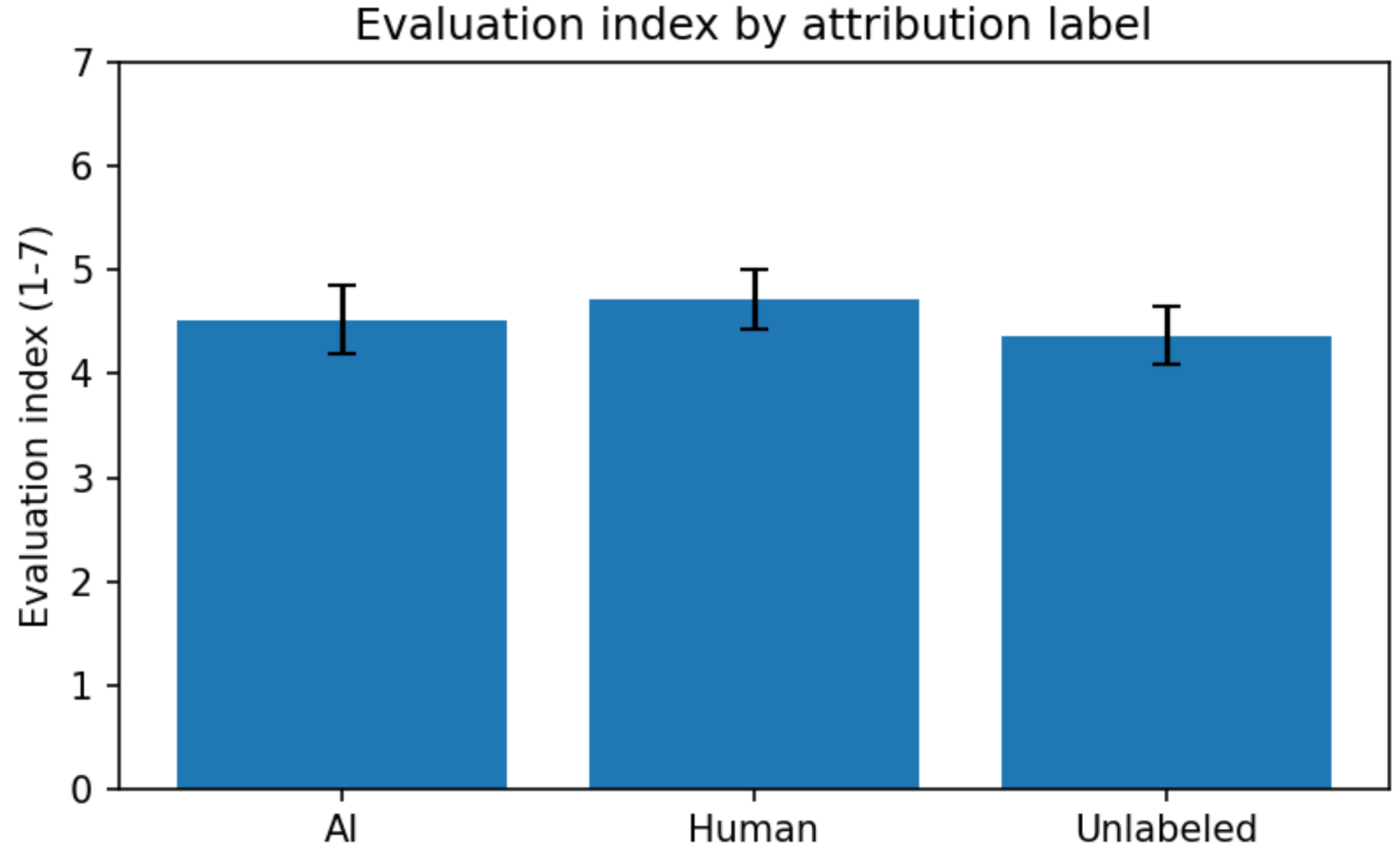

**Figure 2. Enjoyment vs. ln(perceived minutes + 1), by attribution label, with overall regression line.**

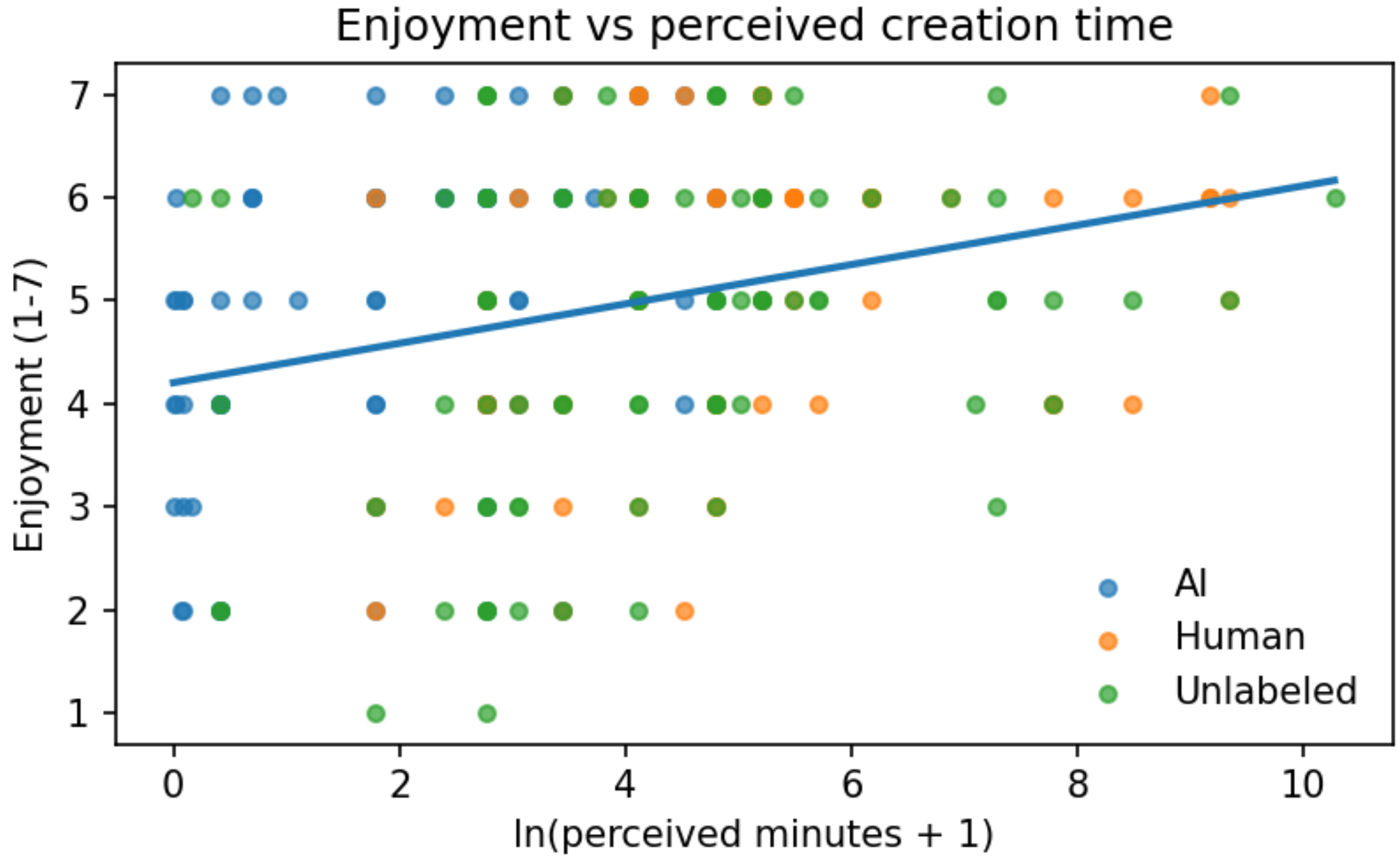